\documentclass[12pt,a4paper,twoside]{article}
\usepackage[T1]{fontenc}
\usepackage{amsmath,amsfonts,amssymb,amsthm,mathabx}
\usepackage{cite}
\usepackage{graphicx}
\usepackage{stmaryrd}
\usepackage{hyperref}
\usepackage{pdfsync}

\newcommand{\be}{\begin{equation}}
\newcommand{\ee}{\end{equation}}
\newcommand{\bea}{\begin{eqnarray}}
\newcommand{\eea}{\end{eqnarray}}

\newcommand{\phii}{\phi}
\newcommand{\BMS}{\mathrm{BMS}_3}

\newcommand{\Diff}{\mathrm{Diff}^+(S^1)}

\newcommand{\hatVect}{\widehat{\mathrm{Vect}}(S^1)}

\newcommand{\hatbms}{\widehat{\mathfrak{bms}}_3}
\newcommand{\hatBMS}{\widehat{\mathrm{BMS}}_3}
\newcommand{\ket}{\rangle}
\newcommand{\bra}{\langle}
\newcommand{\calG}{\mathfrak{g}}

\newcommand{\calB}{{\cal B}}

\newcommand{\calE}{{\cal E}}

\newcommand{\calH}{{\cal H}}

\newcommand{\calO}{{\cal O}}
\newcommand{\calR}{{\cal R}}

\newcommand{\calT}{{\cal T}}
\newcommand{\calW}{{\cal W}}
\newcommand{\Ob}{{\cal O}_p}

\newcommand{\Ad}{\mathrm{Ad}}
\newcommand{\ad}{\mathrm{ad}}
\newcommand{\SL}{\mathrm{SL}}
\renewcommand{\sl}{\mathfrak{sl}}

\newcommand{\RR}{\mathbb{R}}
\newcommand{\CC}{\mathbb{C}}

\newcommand{\GG}{G\ltimes_{\mathrm{Ad}}\mathfrak{g}_{\text{ab}}}

\def\d_Vphi{\mathrm{d}_V\hspace{-0.06em}\phi}
\def\d_Vphibar{\mathrm{d}_V\hspace{-0.06em}\bar\phi}
\def\d_Vxi{\mathrm{d}_V\hspace{-0.06em}\xi}

\def\be{\begin{eqnarray}}
\def\ee{\end{eqnarray}}
\def\beann{\begin{eqnarray*}}
\def\eeann{\end{eqnarray*}}
\def\beq{\begin{equation}}
\def\eeq{\end{equation}}
\def\ba{\begin{array}}
\def\ea{\end{array}}
\def\ben{\begin{enumerate}}
\def\een{\end{enumerate}}
\def\bea{\begin{eqnarray}}
\def\eea{\end{eqnarray}}

\def\5{\bar }
\def\6{\partial }
\def\7{\hat }
\def\4{\tilde }
\advance\textheight\topskip
\renewcommand{\tilde}{\widetilde}
\renewcommand{\hat}{\widehat}

\renewcommand{\simeq}{\cong}

\renewcommand{\d}{\partial}

\renewcommand{\geq}{\,{\geqslant}\,}
\renewcommand{\leq}{\,{\leqslant}\,}

\newcommand{\binner}[2]{%
  {\langle}\kern-4.15pt{\langle}#1{,}\,#2{\rangle}\kern-4.15pt{\rangle}}

\newcommand{\ffrac}[2]{\raisebox{.5pt}%
  {\footnotesize$\displaystyle\frac{#1}{#2}$}\kern1pt}

\newcommand{\ZZ}{\mathbb{Z}}

\def\cI{\mathcal{I}}

\numberwithin{equation}{section} \makeatletter
\@addtoreset{equation}{section}

\DeclareFontFamily{OT1}{rsfs}{} \DeclareFontShape{OT1}{rsfs}{m}{n}{
<-7> rsfs5 <7-10> rsfs7 <10-> rsfs10}{}
\DeclareMathAlphabet{\mycal}{OT1}{rsfs}{m}{n}

\begin{document}

\title{Notes on the BMS group in three dimensions:\\
II.~Coadjoint representation}

\author{Glenn Barnich and Blagoje Oblak}

\date{}

\def\mytitle{Notes on the BMS group in three dimensions:\\
II.~Coadjoint representation}

\pagestyle{myheadings} \markboth{\textsc{\small G.~Barnich,
    B.~Oblak}}{%
  \textsc{\small Coadjoint representation of BMS$_3$}}

\addtolength{\headsep}{4pt}


\begin{centering}

  \vspace{1cm}

  \textbf{\Large{\mytitle}}


  \vspace{1.5cm}

  {\large Glenn Barnich$^a$ and Blagoje Oblak$^{b}$}

\vspace{.5cm}

\begin{minipage}{.9\textwidth}\small \it  \begin{center}
   Physique Th\'eorique et Math\'ematique \\ Universit\'e Libre de
   Bruxelles and International Solvay Institutes \\ Campus
   Plaine C.P. 231, B-1050 Bruxelles, Belgium
 \end{center}
\end{minipage}

\end{centering}

\vspace{1cm}

\begin{center}
  \begin{minipage}{.9\textwidth}
    \textsc{Abstract}. The coadjoint representation of the BMS$_3$
    group, which governs the covariant phase space of
    three-dimensional asymptotically flat gravity, is investigated. In
    particular, we classify coadjoint BMS$_3$ orbits and show that
    intrinsic angular momentum is free of supertranslation
    ambiguities. Finally, we discuss the link with induced
    representations upon geometric quantization.
  \end{minipage}
\end{center}

\vfill

\noindent
\mbox{}
\raisebox{-3\baselineskip}{%
  \parbox{\textwidth}{\mbox{}\hrulefill\\[-4pt]}} {\scriptsize$^a$
  E-mail: gbarnich@ulb.ac.be\\ $^b$ Research Fellow of the Fund for
  Scientific Research-FNRS Belgium. Email: boblak@ulb.ac.be}

\thispagestyle{empty}
\newpage

\begin{small}
{\addtolength{\parskip}{-2pt}
 \tableofcontents}
\end{small}
\thispagestyle{empty}
\newpage

\section{Introduction}
\label{sec:introduction}

Coadjoint orbits of semi-direct product groups with an Abelian factor
are well-understood \cite{Rawnsley1975,VictorGuillemin16276,%
Li1993a,Baguis1998,Ali2000}.  Their classification involves the same
little groups that appear in the context of ``induced
representations'', that is, the construction of unitary irreducible
representations of the full group from those of the little groups.

The purpose of the present paper is to apply this classification to
the centrally extended $\hatBMS$ group and elaborate on its relation
with asymptotically flat solutions to Einstein's equations in three
dimensions. Indeed, it follows from the considerations in
\cite{Barnich:2006avcorr,Barnich:2010eb,Barnich:2012rz} that the
reduced phase space of three-dimensional
asymptotically flat gravity at null infinity coincides with the
coadjoint representation of $\hatBMS$ at fixed central charges
$c_1=0, c_2=3/G$. As a consequence, this solution space consists of
coadjoint orbits of $\hatBMS$. There is thus a close relation between
classical gravitational solutions and unitary irreducible
representations of their symmetry group, or ``$\BMS$ particles'' in the
terminology of \cite{Barnich:2014kra}. Our objective here is to make
this relation precise.

The plan of the paper is the following. We start, in section
\ref{sec:remarks-relat-coadj}, by reviewing the coadjoint
representation of semi-direct product groups and the classification of
their coadjoint orbits. As an application, the case of the Poincar\'e
group in three dimensions is briefly discussed. Section
\ref{sec:co-adjo-repr} is devoted to a description of the coadjoint
representation of $\hatBMS$ and its relation to three-dimensional
asymptotically flat spacetimes at null infinity. In particular, the
full understanding of the coadjoint orbits is used to complete the
positive energy theorem for asymptotically flat three-dimensional
spacetimes \cite{Barnich:2014zoa} by a discussion of angular momentum
in this context. We end in Section \ref{sec:quantization} by
discussing the link between geometric quantization and induced
representations following
\cite{Rawnsley1975,VictorGuillemin16276,Li1993a,Baguis1998}, and apply
these considerations to the case of the $\hatBMS$ group.

Throughout this work, we will use notations, conventions and results
of \cite{Barnich:2014kra}, except for the fact that the dual of the
action involved in the semi-direct product, $\sigma^*$, will always be
written explicitly. 

Specific coadjoints orbits of related (conformal)
Carroll groups have recently also been discussed in
\cite{Duval:2014uoa,Duval:2014lpa}.

\section{Coadjoint orbits of semi-direct products}
\label{sec:remarks-relat-coadj}

In this section we consider a semi-direct product group
$H=G\ltimes_{\sigma}A$, with $G$ a Lie group, $A$ an Abelian vector
group, and $\sigma$ a smooth representation of $G$ in $A$. For
simplicity, we will restrict the discussion here to finite-dimensional
Lie groups.

\subsection{Coadjoint representation}

The Lie algebra of $H$ is $\mathfrak{h}=\mathfrak{g}\oright_{\Sigma}A$
and the adjoint action of $H$ reads
\begin{eqnarray} 
  \Ad_{(f,\alpha)}(X,\beta) = \left( \Ad_f
  X,\sigma_f\beta-\Sigma_{\Ad_fX}\alpha\right)
  \quad\forall\;(f,\alpha)\in H,\;\;\forall\;(X,\beta)\in\mathfrak{h}.
  \nonumber
\end{eqnarray} 
The dual space of $\mathfrak{h}$ is
$\mathfrak{h}^*=\mathfrak{g}^*\oplus A^*$, whose elements, denoted as
$(j,p)$ with $j\in\calG^*$ and $p\in A^*$, are paired with
$\mathfrak{h}$ according to\footnote{We use the same notation
$\bra.,.\ket$ for the pairings of $\mathfrak{h}^*$ with
$\mathfrak{h}$, of $\calG^*$ with $\calG$ and of $A^*$ with $A$.}
\begin{eqnarray} 
  \bra(j,p),(X,\alpha)\ket = \bra j,X\ket+\bra
  p,\alpha\ket.  \nonumber
\end{eqnarray} 
Writing down the coadjoint action of $H$ requires some
additional notation \cite{Rawnsley1975,Baguis1998}: a bilinear ``cross''
product $\times:A\times A^*\rightarrow\calG^*:(\alpha,p)\mapsto
\alpha\times p$ is defined by 
\begin{eqnarray} 
  \bra\alpha\times
  p,X\ket := \bra p,\Sigma_X\alpha\ket \quad\forall\;X\in\calG.
\label{wedge}
\end{eqnarray}
The notation is justified by the fact that, when $H$ is the Euclidean
group in three dimensions, $\alpha\times p$ can be identified with the
usual cross product in $\RR^3$. With this definition, the coadjoint
action of $H$ is given by
\begin{eqnarray}
  \Ad^*_{(f,\alpha)}(j,p)  =  \left(\Ad^*_f j+\alpha\times\sigma^*_f
  p, \sigma^*_fp \right),
\label{coadH}
\end{eqnarray} 
where $\sigma^*$ denotes the dual representation associated with
$\sigma$, while the $\Ad^*$'s on the right-hand side denote the
coadjoint representation of $G$. More generally, the notation $\Ad^*$
will be reserved for the coadjoint representations of both $H$ and
$G$, the subscript indicating which group we are working with.

A special class of such semi-direct products consists of groups of the
form $H=G\ltimes_{\Ad}\mathfrak g_{\rm ab}$, with $A=\calG_{\rm ab}$
seen as an Abelian vector group. In this case,
$\alpha\times p=\ad^*_{\alpha} p$, with an obvious abuse of notation
consisting in identifying elements of $\mathfrak{g}_{\rm ab}$,
respectively $\mathfrak{g}^*_{\rm ab}$, with the corresponding
elements of $\mathfrak{g}$ and $\mathfrak{g}^*$. We use the
index ``${\rm ab}$'' to distinguish the dual space of the Abelian
algebra from that of the non-Abelian one. This class includes the
Euclidean group in three dimensions, the Poincar\'e group in three
dimensions and the $\hatBMS$ group. The dual of $\mathfrak{h}$ then
becomes $\mathfrak{h}^*=\mathfrak{g}^*\oplus\mathfrak{g}^*_{\rm ab}$
and the coadjoint action (\ref{coadH}) reduces to
\begin{equation*}
  \Ad^*_{(f,\alpha)}(j,p)=\big(\Ad^*_fj+\ad^*_\alpha\Ad^*_fp,
  \Ad^*_fp\big).
\end{equation*}

\subsection{Coadjoint orbits}
\label{coadOrb}

The coadjoint orbit of $(j,p)\in\mathfrak{h}^*$ is the set 
\begin{eqnarray}
  \calW_{(j,p)} = \left\{ \Ad^*_{(f,\alpha)}(j,p)|(f,\alpha)\in
  H \right\} \subset\mathfrak{h}^*, \nonumber 
\end{eqnarray} 
with $\Ad^*_{(f,\alpha)}(j,p)$ given by (\ref{coadH}). The goal is to
classify the coadjoint orbits of $H$, assuming that the orbits and
little groups 
\begin{eqnarray} 
  \calO_p = \left\{ \sigma^*_fp|f\in G \right\}\simeq G/G_p, 
\quad G_p = \left\{ f\in
  G|\sigma^*_fp=p \right\} \nonumber 
\end{eqnarray} 
of the action $\sigma^*$ are known. These are the orbits and little
groups that play a key role for induced representations of semi-direct
product groups. From the second half of the right-hand side of
(\ref{coadH}), involving only $\sigma^*_f p$, it follows that each
coadjoint orbit $\calW_{(j,p)}$ is a fibre bundle over the orbit
$\calO_p$, the fibre above $q=\sigma^*_f p$ being the
set 
\begin{eqnarray}
  \left\{ \left(\Ad^*_g \Ad^*_f j+\alpha\times
  q,q\right)|g\in G_q,\alpha\in A \right\}
  \subset\mathfrak{h}^*.  \nonumber 
\end{eqnarray} 
It remains to understand the geometry of these fibres and the relation
between fibres at different points.

\subsubsection*{Preliminary: orbits passing through $j=0$}

Consider the first half of the right-hand side of (\ref{coadH}),
\begin{eqnarray}
\Ad^*_f j+\alpha\times\sigma^*_f p,
\label{rhs}
\end{eqnarray} 
and take $j=0$ for now. Then, keeping $q=\sigma^*_f p$ fixed, the set
spanned by elements of the form (\ref{rhs}) is 
\begin{eqnarray}
  {\rm Span}_A:=\left\{\alpha\times q| \alpha\in A\right\}\subset\calG^*.
\label{awedgeq}
\end{eqnarray} 
Now, the tangent space of $\calO_p$ at $q$ can be identified with
the space of ``infinitesimal displacements'' of $q$: 
\begin{eqnarray} 
  T_q\calO_p =
  \left\{ \Sigma^*_X q| X\in\calG \right\}\subset A^*.
\label{tSpace}
\end{eqnarray} 
Note that $\Sigma^*_X q=0$ iff $X$ belongs to the Lie algebra
$\calG_q$ of the little group $G_q$, so that the tangent space
(\ref{tSpace}) is isomorphic to the coset space $\calG/\calG_q$. It
follows that the cotangent space $T_q^*\calO_p$ at $q$ is the
annihilator of $\calG_q$ in $\calG^*$, that is,
\begin{eqnarray} 
T_q^*\calO_p=\calG_q^0:= \left\{
    j\in\calG^*|\bra j,X\ket=0\;\;\forall\,X\in\calG_q \right\}
  \subset\calG^*.  \nonumber 
\end{eqnarray} 
In turn, the latter space coincides with the set (\ref{awedgeq}): 
\begin{eqnarray}
  T_q^*\calO_p=\calG_q^0={\rm Span}_A.
\nonumber
\end{eqnarray} 


\begin{minipage}{.90\textwidth}\footnotesize

  Indeed, for all $X\in\calG_q$, one has
  $\bra\alpha\times q,X\ket=\bra q,\Sigma_X \alpha\ket=-\bra\Sigma^*_X
  q,\alpha\ket=0$,
  so $\alpha\times q\in\calG_q^0=T^*_q\calO_p$ for all $\alpha\in A$.
  Conversely, any element of $\calG_q^0$ can be written as
  $\alpha\times q$ for some $\alpha\in A$. To see this, consider the
  linear map
  $\tau_q:A\rightarrow\calG_q^0:\alpha\mapsto\alpha\times q$.  The
  image of $\tau_q$ has dimension
  $\text{dim}\, A-\text{dim}\,\text{Ker}(\tau_q)$. Since
  $\text{Ker}(\tau_q) = \left\{ \alpha\in
    A|\bra\Sigma^*_Xq,\alpha\ket=0\;\;\forall\;[X]\in\calG/\calG_q
  \right\}$,
  elements of $\text{Ker}(\tau_q)$ are elements of $A$ constrained by
  $\text{dim}\, \calG-\text{dim}\, \calG_q$ independent conditions.
  This implies that
  $\text{dim}\,\text{Ker}(\tau_q)
  =\text{dim}\,A-\text{dim}\,\calG+\text{dim}\,\calG_q$,
  so that
  $\text{dim}\,\text{Im}(\tau_q)
  =\text{dim}\,\calG-\text{dim}\,\calG_q =\text{dim}\,\calG_q^0$.
  It follows that $\tau_q$ is surjective. \hfill $\blacksquare$

\end{minipage}

\vspace*{.25cm}

\noindent From this we conclude, more generally, that the orbit
passing through $j=0$ is the cotangent bundle of $\calO_p$:
\begin{eqnarray} 
  \calW_{(0,p)}  = \left\{\left(\alpha\times\sigma^*_f p,\sigma^*_f p
  \right)|(f,\alpha)\in H\right\} =
  \bigsqcup_{q\in\calO_p}T_q^*\calO_p = 
  T^*\calO_p \subset\mathfrak{h}^*. \nonumber 
\end{eqnarray}

\subsubsection*{General case}

It remains to understand the role of $j$ in (\ref{rhs}). Let us
therefore fix some $(j,p)\in\mathfrak{h}^*$ and focus for now on
elements $f$ belonging to the little group $G_p$, so that
$\sigma^*_f(p)=p$. With this restriction, the set of points reached by
the coadjoint action of $H$ on $(j,p)$ is
\begin{eqnarray}
\left\{\left(\Ad^*_f j+\alpha\times p,p\right)| f\in
  G_p,\alpha\in A\right\}\subset\mathfrak{h}^*,
\label{blah}
\end{eqnarray}
where in general $\Ad^*_f(j)\neq j$ because the little group $G_p$
need not be included in the stabilizer of $j$ for the coadjoint action
of $G$. Now, it follows from (\ref{wedge}) that
\begin{eqnarray}
\Ad^*_f(\alpha\times p)=\sigma_f \alpha \times\sigma^*_f p.
\nonumber
\end{eqnarray}
Together with the requirement that $f$ belongs to the little group at
$p$, this property allows us to rewrite the set (\ref{blah}) as
\begin{eqnarray}
\left\{
\left(
\Ad^*_f\left(j+\beta\times p\right),p
\right)|f\in G_p,\;\beta\in A
\right\}.
\label{blahBis}
\end{eqnarray}
Hence, in particular, translations along
$\beta$ allow one to modify at will all components of $j$ that point
along directions in the annihilator $\calG_p^0$. The only piece of $j$
that is left unchanged by the action of translations is its restriction
$j_p:=j|_{\mathfrak g_p}$ to $\calG_p$, so the set (\ref{blahBis})
can be rewritten as
\begin{eqnarray}
  \underbrace{\left\{\Ad^*_f j_p|f\in
  G_p\right\}}_{\displaystyle\calW_{j_p}}
  \times
  \underbrace{\left\{\alpha
\times p|\alpha\in A\right\}}_{\displaystyle T_p^*\calO_p},
  \quad\quad
\label{blahTris}
\end{eqnarray}
where $\calW_{j_p}\subset \calG_p^*$ denotes the coadjoint orbit of
$j_p\in\calG_p^*$ under the little group $G_p$. Thus, when
$\calW_{(j,p)}$ is seen as a fibre bundle over $\calO_p$, the fibre
above $p$ is the product (\ref{blahTris}) of the cotangent space of
$\calO_p$ at $p$ with the coadjoint orbit of the projection $j_p$ of
$j$ under the action of the little group of $p$.

The same construction would hold at any other point $q$ on $\calO_p$,
except that the relevant little group would be $G_q$. Thus, the fibre
above any point $q=\sigma^*_f p \in\calO_p$ is a product of the
cotangent space of $\calO_p$ at $q$ with the $G_q$-coadjoint orbit
$\calW_{(\Ad^*_fj)_q}$, where $(\Ad^*_fj)_q$ denotes the restriction
of $\Ad^*_fj$ to $\calG_q$. But little groups at different points of
$\calO_p$ are isomorphic: if one chooses a group element $g_q\in G$
such that $\sigma^*_{g_q}(p)=q$, then $G_q=g_q\cdot G_p\cdot g_q^{-1}$
and $\calG_q=\Ad_{g_q}\calG_p$. Therefore, $\calW_{(\Ad^*_fj)_q}$ is
diffeomorphic to $\calW_{j_p}$ for any $q=\sigma^*_f p\in\calO_p$; the
relation between the fibres above $q$ and $p$ is given by the
coadjoint action of $H$.

\subsubsection*{Classification of coadjoint orbits of $H$}

The conclusion of the last paragraph can be used to classify the
orbits of $H$. The {\it bundle of little group orbits} associated with
$(j,p)\in\mathfrak{h}^*$ is defined as
\begin{eqnarray}
  \calB_{(j_p,p)} := \left\{ \left. \left(
  \left(\Ad^*_f j \right)_{\sigma^*_f p},\sigma^*_f p
  \right)\right |f\in G  \right\}. 
\label{bundleLGOBis}
\end{eqnarray}
According to the discussion of the previous paragraph,
$\calB_{(j_p,p)}$ is really the same as $\calW_{(j,p)}$, except that
the cotangent spaces at each point of $\calO_p$ are ``neglected''. The
bundle of little group orbits is thus a fibre bundle over $\calO_p$,
the fibre $F_q$ at $q\in\calO_p$ being a coadjoint orbit of the
corresponding little group $G_q$. The relation between fibres at
different points of $\calO_p$ is given by the coadjoint action of $H$,
or explicitly,
\begin{eqnarray}
(k,q)\in F_q
\quad\text{iff}\quad
\exists\,f\in G\text{ such that }
k=\left(\Ad^*_f j\right)_q\text{ and }q=\sigma^*_f p.
\nonumber
\end{eqnarray}

Conversely, suppose that two elements $p\in A^*$ and $j_0\in\calG_p^*$
are given. The group $G$ can be seen as a principal $G_p$-bundle over
$\calO_p$, equipped with a natural $G_p$-action by multiplication from
the left in each fibre. In addition, $G_p$ acts on the coadjoint orbit
$\calW_{j_0}$, so one can define an action of $G_p$ on
$G\times\calW_{j_0}$ by
\begin{eqnarray}
(f,k)\in G\times\calW_{j_0}\stackrel{g\in G_p}
{\longmapsto}\left(g\cdot f,\Ad^*_g(k)\right).
\nonumber
\end{eqnarray}
The corresponding bundle of little group orbits $\calB_{(j_0,p)}$ is
defined as 
\begin{eqnarray}
\calB_{(j_0,p)}:=\left(G\times\calW_{j_0}\right)/G_p.
\label{bundleLGO}
\end{eqnarray}
Thus, from each coadjoint orbit of $H$, one can build a bundle of
little group orbits (\ref{bundleLGOBis}); conversely, from each bundle
of little group orbits as defined in (\ref{bundleLGO}), one can build
a coadjoint orbit of $H$ by choosing any $j\in\calG^*$ such that
$j_p=j_0$ and taking the orbit $\calW_{(j,p)}$. In other words, the
classification of coadjoint orbits of $H$ is equivalent to the
classification of bundles of little group orbits
\cite{Rawnsley1975,Baguis1998}. 

This yields the complete picture of coadjoint orbits of $H$: each
coadjoint orbit $\calW_{(j,p)}$ is a fibre bundle over $\calO_p$, the
fibre above $q\in\calO_p$ being a product of $T^*_q\calO_p$ with a
coadjoint orbit of the corresponding little group $G_q$. Equivalently,
$\calW_{(j,p)}$ is a fibre bundle over $T^*\calO_p$, the fibre above
$(q,\alpha\times q)\in T^*\calO_p$ being a coadjoint orbit of
$G_q$. To exhaust all coadjoint orbits of $H$, one proceeds as
follows:
\begin{enumerate}
\item Pick an element $p\in A^*$ and compute its orbit $\calO_p$ under
  the action $\sigma^*$ of $G$;
\item Find the corresponding little group $G_p$;
\item Pick $j_p\in\calG_p^*$ and compute its coadjoint orbit under the
  action of $G_p$.
\end{enumerate}
The set of all orbits $\calO_p$ and of all coadjoint orbits of the
corresponding little groups classifies the coadjoint orbits of
$H$. Put differently, suppose one has classified the following
objects:
\begin{enumerate}
\item The orbits of $G$ for the action $\sigma^*$, with orbit
  representatives $p_{\lambda}\in A^*$ and corresponding little groups
  $G_{\lambda}$, $\lambda\in{\cal I}$ being some index such that
  $\calO_{p_{\lambda}}$ and $\calO_{p_{\lambda'}}$ are disjoint
  whenever $\lambda\neq\lambda'$;
\item The coadjoint orbits of each $G_{\lambda}$, with orbit
  representatives $j_{\lambda,\mu}\in\calG_{\lambda}^*$,
  $\mu\in{\cal J}_{\lambda}$ being some index such that
  $\calW_{j_{\lambda,\mu}}$ and $\calW_{j_{\lambda,\mu'}}$ are
  disjoint whenever $\mu\neq\mu'$.
\end{enumerate}
Then, the set
\begin{equation*}
  \left.\big\{\left(j_{\lambda,\mu},p_{\lambda}
    \right)\right|\lambda\in\cal I,\mu\in\cal J_{\lambda}\big\} \subset
  \mathfrak{h}^*
\end{equation*}
forms a complete set of representatives for the collection of disjoint
coadjoint orbits $\calW_{\left(j_{\lambda,\mu},p_{\lambda}\right)}$ of
$H$. The (possibly continuous) indices $\lambda,\mu$ label the orbits
uniquely.

\subsection{Poincar\'e group in three dimensions}
\label{subsecPoincare}

The double cover of the Poincar\'e group in three dimensions is
\begin{eqnarray}
\SL(2,\RR)\ltimes_{\Ad}\mathfrak{sl}(2,\RR)_{\text{ab}},
\label{poinc}
\end{eqnarray}
with $\SL(2,\RR)$ the double cover of the connected Lorentz group in
three dimensions, and $\sl(2,\RR)_{\text{ab}}$ isomorphic to the
Abelian group of translations. The dual of the Poincar\'e algebra
consists of pairs $(j,p)$, where both $j$ and $p$ belong to
$\sl(2,\RR)^*$. One may refer to $p$ as a momentum vector and to $j$
as an angular momentum vector. The projection $j_p$ of $j$ on
$\calG_p^*$ is the classical analogue of intrinsic spin, while the
components of $j$ that can be varied through translations
$\alpha\times p$ represent orbital angular momentum.

The coadjoint orbits of (\ref{poinc}) are classified by the general
results of subsection \ref{coadOrb}. Let therefore
$\{p_{\lambda}|\lambda\in\cI\}$ be an exhaustive set of
representatives for the coadjoint orbits of $\SL(2,\RR)$, see
e.g.~\cite{Binegar1982,Radu,Witten:1987ty}. For example, take
$\cI=\RR\cup i\RR^+_0\cup\{e^{\pm i\pi/4}\}$ with $\lambda=0$
corresponding to the vanishing momentum;
$p_0=\lambda\in\RR^\pm_0, p_i=0$ corresponding to positive (negative)
energy and positive mass squared; $p_0=0$,
$p_1=0,p_2=i\lambda$, $\lambda\in i\RR^+_0$, corresponding to
negative mass squared; and
$p_2+ip_0=e^{\pm i\pi/4}$, $p_1=0$ corresponding to massless momenta
with positive or negative energy.

Whenever $p_{\lambda}\neq0$, the little group $G_{\lambda}$ is Abelian
and one-dimensional, so that $\mathfrak{g}_{\lambda}^*\simeq\RR$ and
the coadjoint action of $G_{\lambda}$ is trivial. The index $\mu$ in
the set $\{j_{\lambda,\mu}\}$ then runs over all real values,
labelling the component of $j_{\lambda,\mu}$ along the direction
$\mathfrak{g}_{\lambda}^*$ in $\mathfrak{sl}(2,\RR)^*$. We will denote
this component by $s$, for ``spin''. Hence, whenever
$p_{\lambda}\neq0$, the coadjoint orbit of
$(j_{\lambda,\mu},p_{\lambda})$ under the Poincar\'e group is
diffeomorphic to the cotangent bundle
$T^*\calO_{p_{\lambda}}$. However, two such orbits having the same
index $\lambda$, but different indices $\mu$ (i.e.~different spins $s$
of $j_{\lambda,\mu}$ along $\mathfrak{g}_{\lambda}^*$), are
disjoint. The only orbits left are those containing $p=0$. These are
all of the form $\calW_j\times\{0\}\simeq\calW_j$, where $\calW_j$ is
the coadjoint orbit of $j\in\mathfrak{sl}(2,\RR)^*$ under
$\SL(2,\RR)$.

\section{Coadjoint orbits of BMS$_3$}
\label{sec:co-adjo-repr}

As before, we use the notations and conventions of
\cite{Barnich:2014kra}, to which we refer for a definition of the
$\BMS$ group (and its central extension) and the construction of its
induced representations. The purpose of this section is to classify
the coadjoint orbits of the $\BMS$ group and to establish the link of
this classification with three-dimensional gravity. In particular,
angular momentum is studied in some detail.

\subsection{Generalities on the $\BMS$ group}
\label{sec:generalities}

The centrally extended $\hatBMS$ group is of the form
$G\ltimes_{\Ad}\calG_{\text{ab}}$ with $G$ the universal cover of the
Virasoro group. The dual of the $\hatbms$ algebra is the space
$\hatVect^*\oplus\hatVect^*_{\text{ab}}$, whose elements are
quadruples $(j,ic_1;p,ic_2)$, where $c_1$ and $c_2$ are central
charges while the supermomentum $p$ and the angular supermomentum $j$
are quadratic differentials on the circle. The pairing with elements
of $\hatbms$ is explicitly given by
\begin{equation}
  \nonumber
  \langle (j,ic_1;p,ic_2), (X,-ia;\alpha,-ib)\rangle
  =\int^{2\pi}_0d\phii\, 
  \left(j(\phi)X(\phi)+p(\phi)\alpha(\phi)\right)
  +c_1a+c_2b.
\end{equation}
Accordingly, the coadjoint action of the $\hatbms$ algebra is
\begin{equation}
  \label{eq:34}
\ad^*_{(X,\alpha)}(j,ic_1;p,ic_2)=(\delta j\,d\phii^2,0;\delta p\;d\phii^2,0),
\end{equation}
with
\begin{equation}
\label{eq:34bis}
\delta p
=
Xp'+2X'p-\frac{c_2}{24\pi}X''',\quad
\delta j
=
Xj'+2X'j-\frac{c_1}{24\pi}X'''+\alpha p'+2\alpha'p-\frac{c_2}{24\pi}\alpha'''.
\end{equation}
The corresponding\footnote{More precisely,
  \eqref{eq:34}-\eqref{eq:34bis} is the differential of
  \eqref{eq:34a}-\eqref{eq:34abis} up to an overall minus sign.}
coadjoint representation of the $\hatBMS$ group is given by
\begin{equation}
 \label{eq:34a}
 \Ad^*_{(f,\alpha)^{-1}}(j,ic_1;p,ic_2)
 =
 \left({\tilde j}d\phii^2,ic_1;{\tilde p}d\phii^2,ic_2\right),
\end{equation}
where
\begin{equation}
\label{eq:34abis}
\tilde p
=
(f')^2p\circ f- \frac{c_2}{24\pi}S[f],
\quad
\tilde j
=
(f')^2\left[j+\alpha p'+2\alpha'p-\frac{c_2}{24\pi}\alpha'''\right]
\circ f-\frac{c_1}{24\pi}S[f],
\end{equation}
and $S[f]=f'''/f'-\frac{3}{2}(f''/f')^2$ denotes the Schwarzian
derivative of $f$.

\subsection{Coadjoint orbits}
\label{sec:coadjoint-orbits}

As for the Poincar\'e group, the coadjoint orbits of $\hatBMS$ are
classified according to the general results of subsection
\ref{coadOrb}. Due to the structure $G\ltimes\calG_{\text{ab}}$ of
$\hatBMS$, the orbits denoted $\calO_p$ in section
\ref{sec:remarks-relat-coadj} are the well-known coadjoint orbits of
the Virasoro group, see
e.g.~\cite{LazPan,Segal:1981ap,Witten:1987ty,Balog:1997zz,guieu2007}.
Nevertheless, we will keep calling these orbits ``orbits of the action
$\sigma^*$'' in order to distinguish them from the coadjoint orbits of
$\hatBMS$ itself.

For a given $(j,ic_1;p,ic_2)\in\hatbms^*$, the coadjoint orbit
$\calW_{(j,ic_1;p,ic_2)}$ is thus a bundle over the cotangent bundle
of the orbit $\calO_{(p,ic_2)}$, the typical fibre being a coadjoint
orbit of the corresponding little group. In the case at hand, each
$\calO_{(p,ic_2)}$ is a coadjoint orbit of the Virasoro group. Because
the structure of Virasoro coadjoint orbits depends crucially on the
(non-)vanishing of the central charge $c_2$, we will focus first on
the case that is relevant for three-dimensional gravity, namely
$c_2\neq0$.

A generic orbit $\calO_{(p,ic_2)}$ then has a one-dimensional
(Abelian) little group $G_{(p,ic_2)}$, whose coadjoint representation
is trivial. In particular, as in the Poincar\'e group, little
group orbits consist of only one point, specified by the real value
$(s,ic_1)$ of $(j,ic_1)_{(p,ic_2)}$ in $\calG_{(p,ic_2)}^*$; the value
$s$ can again be considered as the classical analogue of spin. Hence,
for generic supermomenta $p$, the coadjoint orbit
$\calW_{(j,ic_1;p,ic_2)}$ is diffeomorphic to the cotangent bundle
$T^*\calO_{(p,ic_2)}$ and is specified by (i) the value of the central
charges $c_1$ and $c_2\neq 0$, (ii) the supermomentum $p$, and (iii)
the spin $s$.

Still working at non-zero $c_2$, we also need to consider
``exceptional'' (non-generic) orbits $\calO_{(p,ic_2)}$ of constant
supermomenta satisfying $p=-n^2c_2/48\pi$ for some positive
integer $n$. For such orbits, the little group is the $n$-fold cover
of $\mathrm{PSL}(2,\RR)$, so, in contrast to generic orbits, the
little group's coadjoint representation is not trivial. The
coadjoint orbit $\calW_{(j,ic_1;p,ic_2)}$ then is a fibre bundle over
$T^*\calO_{(p,ic_2)}$, having a coadjoint
orbit of $\mathrm{PSL}^{(n)}(2,\RR)$ as its typical fibre.

The case of vanishing $c_2$ is more intricate, because then the little
group may have arbitrary dimension (see e.g.~the summary in
\cite{Khesin2009}, section 2.2). We will not consider this situation
in full generality here. Let us only mention one special case: take
$c_2=0$ and $p=0$ and consider the coadjoint orbit
$\calW_{(j,ic_1;0,0)}$, which is diffeomorphic to a coadjoint orbit of
the Virasoro group for central charge $c_1$ and quadratic differential
$j$. The central charge $c_1$ then plays a crucial role, in contrast
to the case $c_2\neq 0$ discussed above. Such orbits are the $\BMS$
analogue of the Poincar\'e orbits $\calW_{(j,0)}$.

\subsection{Covariant phase space of asymptotically flat gravity}
\label{sec:covar-phase-space-1}

\subsubsection*{Preliminary: the AdS case}

The general solution of Einstein's equations in three dimensions with
negative cosmological constant $\Lambda=-1/\ell^2$ and Brown-Henneaux
boundary conditions \cite{Brown:1986nw} can be written as
\cite{Banados1999,Skenderis:1999nb}
\begin{eqnarray}
  ds^2
  =
  \frac{\ell^2}{r^2}dr^2
  -r^2\big(dx^+-\frac{8\pi G\ell}{r^2}L^-dx^-\big)
  \big(dx^--\frac{8\pi G\ell}{r^2}L^+dx^+\big)
\label{solAdS}
\end{eqnarray}
in terms of a radial coordinate $r\in\RR^+$ and light-cone coordinates
$x^{\pm}=t/\ell \pm\phi$ on the cylinder, where $L^+(x^+)$ and
$L^-(x^-)$ are arbitrary, smooth, $2\pi$-periodic functions. Under the
action of conformal transformations of the cylinder at infinity, these
functions transform according to the coadjoint representation of the
Virasoro group with central charges
\begin{eqnarray}
c^{\pm}=3\ell/2G.
\label{BrownHenneaux}
\end{eqnarray}
In this sense, the space of solutions of Einstein gravity on AdS$_3$
coincides with the hyperplane, at fixed central charges
(\ref{BrownHenneaux}), of the dual space of two copies of the Virasoro
algebra. In particular, solutions of AdS$_3$ gravity are classified by
Virasoro coadjoint orbits \cite{Navarro-Salas1999,Nakatsu1999}, and
this classification defines a symplectic foliation of the space of
solutions. From the point of view of the AdS/CFT correspondence, this
property should not appear as a surprise, given that the operator dual
to the bulk metric in AdS/CFT is the energy-momentum tensor, which, in
two-dimensional conformal field theories, transforms precisely under
the coadjoint representation of the Virasoro group.

\subsubsection*{The flat case}

A similar classification can be implemented for asymptotically flat
space-times. Indeed, in BMS coordinates $(r,u,\phi)$, the general
solution of Einstein's equations in three dimensions describing
asymptotically flat spacetimes at null infinity is given by metrics
\cite{Barnich:2010eb}
\begin{equation}
  \nonumber
  ds^2=\Theta du^2-2dudr+\left(2\Xi+u\Theta'\right)dud\phii
  +r^2d\phii^2 
\end{equation}
depending on two arbitrary functions on the circle,
$\Theta=\Theta(\phii)$ and $\Xi=\Xi(\phii)$. Under finite $\BMS$
transformations acting on the cylinder at null infinity, these
functions have been shown \cite{Barnich:2012rz,Barnich:2013yka} to
transform according to the coadjoint representation
\eqref{eq:34}-\eqref{eq:34abis} upon identifying $\Theta=(16\pi G) p$ 
and $\Xi = (8\pi G) j$, with central charges
\begin{equation}
c_1=0,\quad c_2=3/G.\label{eq:28}
\end{equation}
Thus, just as for AdS$_3$ space-times, the space of solutions of
Einstein's equations with suitable flat boundary conditions at null
infinity\footnote{Other boundary boundary conditions are of course
  possible, leading either to a more restricted symmetry and dynamical
  structure \cite{Henneaux:1984ei,Deser:1985qs} or to an enhancement
  with Weyl symmetry \cite{Barnich:2010eb}.}  is a hyperplane, at
fixed central charges (\ref{eq:28}), in the dual space of the
$\hatbms$ algebra. As a consequence, these solutions can be classified
according to $\hatBMS$ coadjoint orbits, and this classification again
splits solution space into disjoint symplectic leaves. For example,
the solution corresponding to $\Theta=-n^2$ and $\Xi=0$ represents
Minkowski spacetime for $n=1$, and a conical excess of $2\pi n$ for
$n>1$. Its coadjoint orbit $\calW_{(0,0;p,ic_2)}$ is diffeomorphic to
the cotangent bundle of the Virasoro orbit
$\Diff/\mathrm{PSL}^{(n)}(2,\RR)$. Other zero-mode solutions (that is,
solutions specified by constant $\Theta$ and $\Xi$ without
$\Theta=-n^2$) represent cosmological solutions, angular defects or
angular excesses, depending on the sign of $\Theta$ and $\Theta+1$
\cite{Deser:1983tn,Ezawa:1992nk,Cornalba:2002fi,Cornalba:2003kd,%
  Barnich:2012aw}; when seen as elements of $\hatbms^*$, their
coadjoint orbits are all diffeomorphic to the cotangent bundle of
$\Diff/S^1$.

\subsubsection*{Surface charge algebra revisited}

The identification of the space of solutions with the coadjoint
representation of the asymptotic symmetry group can be understood from
the expression of the surface charges\footnote{A similar observation
  also holds in the ${\rm AdS}$ case.}: for
$(X,\alpha)\in\mathfrak{bms}_3$, the latter are given by
\cite{Barnich:2010eb}
\begin{equation}
  \label{eq:47}
  Q_{(X,\alpha)}[\Xi,\Theta]
  =  \frac{1}{16\pi G}\int_0^{2\pi}d\phii\,\left[2\Xi(\phi)X(\phi)
+\Theta(\phi)\alpha(\phi)\right]=
\bra (j,p),(X,\alpha)\ket,
\end{equation}
when writing $\Theta=16\pi G p$ and $\Xi=8\pi Gj$. In turn, this gives
a physical interpretation to the coadjoint vectors $p$ and $j$ of
$\hatbms^*$ in the present context: they represent Bondi mass and
angular momentum aspects.

For any Lie group $G$ with Lie algebra $\calG$, there is a natural
Poisson bracket on $\calG^*$, defined, for any pair of smooth
functions $\Phi,\Psi:\calG^*\rightarrow\RR$, as
\begin{eqnarray}
\left\{\Phi,\Psi\right\}(j)
:=
\bra j,\left[d\Phi_j,d\Psi_j\right]\ket
\quad\forall\;j\in\calG^*.
\label{KKSbra}
\end{eqnarray}
Alternatively, in terms of coordinates $x_a$ on $\calG^*$, 
\begin{equation}
  \label{eq:1}
  \{x_a,x_b\}=f^c_{ab}x_c.
\end{equation}

For the linear maps $\langle
(j,ic_1;p,ic_2),(X,-ia;\alpha,-ib)\rangle$ parametrized by Lie algebra
elements $(X,-ia;\alpha,-ib)$, the Poisson
bracket (\ref{KKSbra}) yields
\begin{multline}
  \nonumber \left\{ \langle (j,ic_1;p,ic_2),(X_1,-ia_1;\alpha_1,-ib_1)\rangle,
    \langle (j,ic_1;p,ic_2),(X_2,-ia_2;\alpha_2,-ib_2)\rangle\right\} =\\=
  \langle (j,ic_1;p,ic_2),\left[(X_1,-ia_1;\alpha_1,-ib_1),
    (X_2,-ia_2;\alpha_2,-ib_2)\right]\rangle.
\end{multline}
When $c_1=0,c_2=3/G$, these brackets correspond precisely to the Dirac
brackets of the surface charges (\ref{eq:47}) as computed in
\cite{Barnich:2006avcorr}. Equivalently, in terms of coordinates
$j(\phi),p(\phi),c_1,c_2$ on $\hatbms^*$, \eqref{eq:1} becomes
\begin{equation}
  \nonumber
  \begin{split}
    &
    \{j(\phii),j(\phii')\}=(j(\phii)+j(\phii'))
\d_\phii\delta(\phii-\phii')
    -\frac{c_1}{24\pi}\d^3_\phii
    \delta(\phii-\phii'), \\
    &
    \{p(\phii),j(\phii')\}=(p(\phii)+p(\phii'))
\d_\phii\delta(\phii-\phii')
    -\frac{c_2}{24\pi}\d^3_\phii \delta(\phii-\phii'), \\
    & \{p(\phii),p(\phii')\}= 0,
  \end{split}
\end{equation}
while brackets involving $c_1,c_2$ vanish. 

\subsection{Energy and angular momentum}
\label{sec:energy-angul-moment}

Classifying the space of solutions of three-dimensional gravity
according their asymptotic symmetries allows one to study properties
of energy and angular momentum, such as boundedness for instance
\cite{Garbarz:2014kaa,Barnich:2014zoa} (see also for
\cite{Raeymaekers:2014kea} for other recent considerations). Both in
the AdS$_3$ and in the flat case, energy turns out to be related to a
zero-mode of a coadjoint vector of the Virasoro group, and the
boundedness properties of this zero-mode are known
\cite{Witten:1987ty,Balog:1997zz}.

In the AdS$_3$ case, with general solution (\ref{solAdS}) and central
charges (\ref{BrownHenneaux}), the only solutions that belong to
Virasoro orbits whose energy is bounded from below are specified by
pairs $(L^+,L^-)$ in which both functions $L^{\pm}$ belong either to
the orbit of a constant $L^{\pm}_{\text{cst}}\geq-c^{\pm}/48\pi$, or
to the orbit of the ``future-directed, massless deformation'' of
$-c^{\pm}/48\pi$. This class of solutions contains, in particular, all
BTZ black holes, but also conical defects and solutions containing
closed time-like curves. On the Virasoro orbits of all such solutions,
the zero-mode $L^{\pm}_0$ of $L^{\pm}$ is bounded from below either by
the value $2\pi L^{\pm}_{\text{cst}}$, or (in the case of the massless
orbit) by the vacuum energy $-c^{\pm}/24$. In particular, since energy
($E$) and angular momentum ($J$) are related to the zero-modes of
$L^{\pm}$ by
\begin{eqnarray}
E=\frac{1}{\ell}\left(L^+_0+L^-_0\right),
\quad
J=L^+_0-L^-_0,
\nonumber
\end{eqnarray}
all solutions of AdS$_3$ gravity that belong to orbits with energy
bounded from below have their angular momentum bounded by
\begin{eqnarray}
|J|\leq \ell E+c^\pm/12.
\label{Jbound}
\end{eqnarray}
Similarly, solutions belonging to the orbit of BTZ black holes, which
correspond to the case where $L^+$ and $L^-$ are positive constants,
all satisfy the cosmic censorship bound $|J|\leq\ell E$.

In the case of asymptotically flat gravity, a natural definition of
total energy and angular momentum is then also simply given by the
zero modes of $p$ and $j$,
\begin{equation}
  E=\int_0^{2\pi}d\phi\, p(\phi),
  \quad
  J=\int_0^{2\pi}d\phi\, j(\phi).
  \label{tot}
\end{equation}  
Following \cite{Misner:1970aa}, chapters 19 and 20, a justification
for this definition goes as follows. Total momentum and total angular
momentum are the surface charges associated with the translation and
the Lorentz Killing vectors of the asymptotically Lorentz frame, for
which $u=x^0-r$, $re^{i\phi}=x^1+ix^2$ (see section 5.3 of
\cite{Barnich:2014kra} for more details). In particular, total energy
and the total rotation vector, which has but one component in three
dimensions, are associated with $\d/\d x^0=\d/\d u$ and $x^1\d/\d
x^2-x^2\d/\d x^1=\d/\d\phi$, which yields \eqref{tot} when used in
\eqref{eq:47}.

The boundedness properties of total energy on coadjoint orbits of
$\hatBMS$ follow from the transformation law
(\ref{eq:34a})-(\ref{eq:34abis}). Because $p$ transforms as a Virasoro
coadjoint vector, without any influence of $j$ or $c_1$, the energy
$E$ has the same boundedness properties as $L^\pm_0$, but with central
charge $c_2$ given in (\ref{eq:28}). Thus, $E$ is bounded from below
on the orbit of $p$ iff this orbit is either that of a constant
$p_{\text{cst}}\geq-c_2/48\pi$, or that of the future-directed,
massless deformation of $-c_2/48\pi$. In particular, all cosmological
solutions and all conical defects in flat space belong to orbits on
which energy is bounded from below. Energy is also bounded from below
on the orbit of Minkowski space (corresponding to $j=0$ and
$p=-c_2/48\pi$), which realizes the minimum value of energy,
$E_{\text{min}}=-c_2/24$.

In order to discuss properties of total angular momentum, we define,
as in the Poincar\'e case, intrinsic angular momentum as total angular
computed in the rest frame. A solution labelled by $(j,p)$ is put in
its rest frame if the supermomentum $p(\phi)$ is brought to a constant
$p_{\text{cst}}$ by using a suitable superrotation. This is of course
not possible on solutions whose $p$ belongs to a Virasoro
coadjoint orbit without constant representative. By integrating on
the circle the piece $\alpha\times p=\ad^*_{\alpha}(p)$ in
\eqref{eq:34abis}, it then follows that:

{\em Intrinsic angular momentum is free from supertranslation
  ambiguities.} 

By contrast, whenever both $p(\phi)$ and $\alpha(\phi)$ are
non-constant on the circle, meaning in particular that the applied
supertranslation is not just a time translation, $\ad^*_{\alpha}(p)$
has a generally non-vanishing zero-mode and contributes to total
angular momentum.

Considering the boundedness properties of the total angular
momentum $J$ as such does not make sense. Whenever $p(\phi)$ is
non-constant on the circle, the value of $J$ can be tuned at will by
acting with supertranslations. In particular, total angular momentum
is unbounded from above and from below on all orbits $\calO_p$. This
should be contrasted with the completely different situation in
AdS$_3$ spacetimes, where separate boundedness properties for the
left- and right-moving energies imply boundedness of total angular
momentum, as in eq. (\ref{Jbound}).

As regards intrinsic angular momentum, the situation is different. By
construction, supertranslations play no role there. In the rest frame,
the only superrotations that are still allowed are those of the little
group $G_{p_{\text{cst}}}$. So for boundedness properties of intrinsic
angular momentum, one needs to study its boundedness properties on
coadjoint orbits of the little group.

Let us illustrate our purposes with two examples. First,
consider the orbit $\calO_p$ of a massive $\BMS$ solution, that is, an
orbit containing a constant supermomentum
$p(\phi)=p_{\text{cst}}>-c_2/48\pi$.  The corresponding little group
$\text{U}(1)$ consists of rigid rotations $f(\phi)=\phi+{\rm cst}$,
and it follows from \eqref{eq:34abis} that intrinsic angular momentum
is unaffected by such superrotations.

Second, consider the vacuum supermomentum vector $p=-c_2/48\pi$, whose
little group is $\text{PSL}(2,\RR)$. It follows from the discussion of
section 5.3 of \cite{Barnich:2014kra} that the coadjoint orbits of
$\text{PSL}(2,\RR)$ are the ``mass hyperboloids'' of the Lorentz group
in three dimensions, represented in a three-dimensional space with
axes $J_0=J$, $J_{\pm 1}=\int_0^{2\pi}d\phi\, j\, e^{\pm i\phi}$.
Depending on the $\text{PSL}(2,\RR)$-orbit, the boundedness properties
of $J$ are very different. The trivial case of the vacuum orbit
$J_0=J_{\pm 1}=0$ brings nothing new, as it is left invariant by the
whole little group $\text{PSL}(2,\RR)$; intrinsic angular momentum
vanishes and total angular momentum is entirely composed of orbital
angular momentum.  By contrast, consider the ``future-directed,
massive orbit'' of $\text{PSL}(2,\RR)$, the upper half of the
two-sheeted hyperboloid. Then, little group transformations act
non-trivially on $J$ but intrinsic angular momentum is bounded from
below by the intersection of the hyperboloid with the $J_0$ axis. It
may, however, take an arbitrarily large value. Similar results hold
for the upper conical orbits of $\text{PSL}(2,\RR)$.  For the
one-sheeted hypeboloid, the ``tachyonic'' orbit, intrinsic angular
momentum is bounded neither from below, nor from above.

\section{Quantization and induced representations}
\label{sec:quantization}

In this section we discuss the relation between classical asymptotically
flat solutions and $\BMS$ particles, i.e.~the link between
coadjoint and induced representations of $\hatBMS$, in terms of
geometric quantization.

\subsection{Geometric quantization for semi-direct products}

\subsubsection*{Generalities on the orbit method}

Let $G$ be a Lie group with Lie algebra $\calG$. The Poisson bracket
(\ref{KKSbra}) defines a symplectic foliation of $\calG^*$, the leaf
through $j\in\calG^*$ being the coadjoint orbit $\calW_j$ of $j$. For
any $k$ belonging to $\calW_j$, the tangent space of $\calW_j$ at $k$
can be identified with the space of ``infinitesimal displacements''
$\ad^*_X(k)$, where $X\in\calG$. The bracket (\ref{KKSbra}) then
induces, on each orbit $\calW_j$, a $G$-invariant symplectic form
$\omega$ given by \cite{kirillov1976elements,Kostant1970,Souriau1970}
\begin{eqnarray}
\omega_k\left(\ad^*_X(k),\ad^*_Y(k)\right):= k\left([X,Y]\right).
\label{KKS}
\end{eqnarray}
Geometric quantization associates a quantum Hilbert space with the
phase space $\calW_j$, proceeding in two steps: prequantization and
polarization.

Prequantization
turns out to be possible provided the symplectic form satisfies the
integrality condition
\begin{eqnarray}
\left[\frac{\omega}{2\pi}\right]\in H^2_{\text{de Rham}}(\calW_j,\ZZ),
\label{integrality}
\end{eqnarray}
in which case there exists a complex line bundle over $\calW_j$,
endowed with a connection whose curvature two-form is
$\omega/2\pi$. The pre-quantum Hilbert space consists of all sections
of this line bundle that are square-integrable with respect to a
Hermitian structure preserved by the connection. This Hilbert space is
then reduced to a smaller subspace by choosing a polarization $-$ an
appropriate subbundle of the complexified tangent bundle of the
symplectic manifold $-$ and restricting both the quantizable
observables and the quantum wavefunctions to be compatible with this
polarization. The idea of the orbit method is that, when a suitable
polarization can be found, geometric quantization should produce an
irreducible unitary representation of the corresponding Lie group
$G$. For compact or solvable Lie groups, this procedure actually
exhausts all irreducible unitary representations; for other Lie
groups, complications may arise, especially if the group is
infinite-dimensional.

\subsubsection*{Semi-direct products}

Suppose we want to quantize a coadjoint orbit $\calW_{(j,p)}$ of the
semi-direct product group $H$. Since the Lie
bracket in $\mathfrak{h}$ reads
\begin{eqnarray}
\left[(X,\beta),(Y,\gamma)\right]
=
\left([X,Y],\Sigma_X\gamma-\Sigma_Y\beta\right),
\nonumber
\end{eqnarray}
the natural symplectic form (\ref{KKS}), evaluated at the point
$\big(\Ad^*_fj+\alpha\times\sigma^*_fp,\sigma^*_fp\big)$ in
$\calW_{(j,p)}$, is given by\footnote{Here we write the argument of
  $\omega^H$ as a pair of elements of the Lie algebra of $H$. This is
  an abuse of notation, being understood that these elements represent
  tangent vectors of $\calW_{(j,p)}$ at the point
  $\left(\Ad^*_f(j)+\alpha\times q,q\right)$ through their coadjoint
  action on that point, as in (\ref{KKS}).}
\begin{eqnarray}
&   & \omega_{\left(\Ad^*_fj+\alpha\times q,q\right)}
\left((X,\beta),(Y,\gamma)\right)=\nonumber\\
\label{omegaH}
& = &
\bra\Ad^*_fj,[X,Y]\ket+\bra\gamma\times q,X\ket
-\bra\beta\times q,Y\ket+\bra\alpha\times q,[X,Y]\ket,
\end{eqnarray}
where we write $q=\sigma^*_fp$ for simplicity. In the three last terms
of this expression, one may recognize the Liouville symplectic form on
the cotangent bundle $T^*\calO_p$, provided $\alpha\times q$ is seen
as an element of $T^*_q\calO_p$. On the other hand, the first term of
(\ref{omegaH}) looks just like the natural symplectic form (\ref{KKS})
on the $G$-coadjoint orbit of $j$ up to the fact that $\Ad^*_fj$ does
{\it not}, generally, belong to $\calG_q^*$. Thus, if we see
$\calW_{(j,p)}$ as a fibre bundle over $T^*\calO_p$ with typical fibre
the coadjoint orbit $\calW_{j_p}$ of the little group, restricting the
symplectic form (\ref{omegaH}) to a fibre gives back the symplectic
form on the little group's coadjoint orbit.

This observation actually follows from a more general result, which
states that the coadjoint orbits of a semi-direct product are obtained
by symplectic induction from the coadjoint orbits of its little
groups, see \cite{Baguis1998,duval1992} for details. This result is of
crucial importance for geometric quantization of
$\calW_{(j,p)}$. Indeed, since the Liouville symplectic form is exact,
it implies that the cohomology class of $\omega$ in (\ref{omegaH})
depends only on the class of the natural symplectic form on the
appropriate coadjoint orbit of the little group. In other words, the
$H$-coadjoint orbit $\calW_{(j,p)}$ is prequantizable iff the
corresponding $G_p$-coadjoint orbit $\calW_{j_p}$ is
prequantizable. Provided a suitable polarization can be found for this
orbit, one obtains a unitary representation $\calR$ of the little
group $G_p$, acting on a complex Hilbert space $\calE$ whose scalar
product we will denote as $(.|.)$. One can then use the representation
$\calR$ of $G_p$ to ``induce'' a representation $\calT$ of $H$. To do
so, one chooses a real polarization for sections of the trivial line
bundle over $T^*\calO_p$, such that polarized sections be functions
$\calO_p\rightarrow\CC$. Provided a $G$-quasi-invariant measure $\mu$
exists on $\calO_p$, the Hilbert space $\calH$ obtained upon
quantization of $\calW_{(j,p)}$ becomes the space of square-integrable
``wavefunctions'' $\Psi:\calO_p\rightarrow\calE$, their scalar product
being
\begin{equation}
\langle\Phi|\Psi\rangle:=\int_{\Ob}d\mu(q)\left(\Phi(q)|\Psi(q)\right).
\label{scalp}
\end{equation}
The action $\calT$ of $H$ on the space of such wavefunctions then
coincides with that of an induced representation
\cite{duval1992,Li1993}. Thus, geometric quantization of the coadjoint
orbits of a semi-direct product group reproduces induced
representations in the sense of Wigner and Mackey.

We should mention, however, that not all induced representations can
be recovered in this way. For instance, if the little group $G_p$ is
disconnected, it may happen that the only representations of $G_p$
available through quantization are those in which the discrete
subgroup of $G_p$ is represented trivially \cite{Li1993}. Furthermore,
the construction may suffer from other complications, related for
instance to the non-existence of a suitable polarization.

\subsection{Gravity and BMS$_3$ particles}
\label{sec:quant-induc-repr}

We can apply the procedure outlined in the previous subsection to
coadjoint orbits of the $\hatBMS$ group, seen as phase spaces equipped
with the natural symplectic form (\ref{omegaH}). As sketched above for
the case of finite-dimensional Lie groups, geometric quantization of
such orbits produces induced representations. Owing to the discussion
of section \ref{sec:co-adjo-repr}, this relation can be rephrased in
terms of solutions of Einstein's equations in an asymptotically flat
space-time in three dimensions: geometric quantization of the orbit
corresponding to a solution labelled by the pair $(j,p)$ produces a
$\BMS$ particle whose supermomenta span the orbit $\calO_p$, and whose
spin is determined by $j_p$. This establishes the link between the
considerations of \cite{Barnich:2014kra} and three-dimensional
gravity.

There is, however, an important subtlety: in writing down the scalar
product (\ref{scalp}), we assumed the existence of a quasi-invariant
measure $\mu$ on $\calO_p$. When $\calO_p$ is a finite-dimensional
manifold, such a measure always exists
\cite{A.O.Barut702}. Furthermore, for semi-direct products of the form
$\GG$, the orbits $\calO_p$ are coadjoint orbits of $G$; they have,
therefore, a symplectic form $\omega$ given by (\ref{KKS}), which can
be used to define an invariant volume form proportional to
$\omega^{d/2}$, where $d$ denotes the dimension of the orbit. But in
the case of $\hatBMS$, the orbits $\calO_p$ are infinite-dimensional
Virasoro orbits, so the question of the existence of a quasi-invariant
measure is much more involved, see
e.g.~\cite{Shavgulidze1997,Bogachev1998,Shimomura2001,Airault2001,Dai2004}. We
will not study this problem here, but hope that it will be settled in
the future.

\section{Conclusion}
\label{sec:conclusion}

In this work we have shown how the classification of coadjoint orbits
of the centrally extended $\hatBMS$ group controls solutions of
asymptotically flat Einstein gravity in three dimensions. Upon
geometric quantization, these orbits yield $\BMS$ particles,
i.e.~induced representations of the $\hatBMS$ group. This brings the
understanding of the relation between group-theoretic aspects of the
$\BMS$ group and flat space gravity to the same level as has been
achieved in the ${\rm AdS}_3$ case in
\cite{Nakatsu1999,Navarro-Salas1999,Garbarz:2014kaa}.

\section*{Acknowledgements}

\addcontentsline{toc}{section}{Acknowledgments}

We are grateful to D.~Pickrell and K.H.~Neeb for helpful comments on
quasi-invariant functional measures. B.O. would also like to thank
J.H.~Rawnsley for useful discussions on geometric quantization for
semi-direct product groups. This work is supported in part by the Fund
for Scientific Research-FNRS (Belgium), by IISN-Belgium, and by
``Communaut\'e fran\c caise de Belgique - Actions de Recherche
Concert\'ees''.



\section*{References}
\addcontentsline{toc}{section}{References}

\renewcommand{\section}[2]{}%

\def\cprime{$'$}
\providecommand{\href}[2]{#2}\begingroup\raggedright\endgroup

\end{document}